\documentclass[english,doublecol,figures]{epl2} 
\usepackage{dcolumn}
\usepackage{amsmath}
\usepackage{amsfonts}
\usepackage{amssymb}
\usepackage{bm}
\usepackage[T1]{fontenc}
\usepackage[latin1]{inputenc}
\usepackage{ae}
\usepackage{graphicx}
\usepackage{epsfig}
\usepackage{psfig}
\usepackage{pstricks,pst-grad}
\usepackage{babel}

\title{Cyclic motion and inversion of surface flow direction in a dense polymer brush under shear}
\shorttitle{Cyclic motion and inversion of surface flow direction in a sheared brush} %Insert here a short version of the title if it exceeds 70 characters

\author{M.~M{\"u}ller\inst{1}\thanks{E-mail:\email{mmueller@theorie.physik.uni-goettingen.de}} \and C.~Pastorino\inst{2}}
\shortauthor{M.~M{\"u}ller and C.~Pastorino}

\institute{                    
  \inst{1} Institut f{\"u}r Theoretische Physik, Georg-August Universit{\"a}t, Friedrich Hund-Platz 1, 37077 G{\"o}ttingen, Germany\\
  \inst{2} Dept.~de F{\'i}sica, Centro At{\'o}mico Constituyentes, CNEA-CONICET, Av.~Gral.~Paz 1499, 
           1650 Pcia.~de Buenos Aires, Argentina
}

\abstract{
Using molecular simulations, we study the properties of a polymer brush in
contact with an explicit solvent under Couette and Poiseuille flow. The solvent
is comprised of chemically identical chains. We present evidence that
individual, unentangled chains in the dense brush exhibit cyclic, tumbling
motion and non-Gaussian fluctuations of the molecular orientations similar to
the behaviour of isolated tethered chains in shear flow. The collective
molecular motion gives rise to an inversion of hydrodynamic flow direction in
the vicinity of the brush-coated surface. Utilising Couette and Poiseuille
flow, we investigate to what extend the effect of a brush-coated surface can be
described  by a Navier slip condition.  }

\begin{document}

\maketitle

\section{Introduction}
Grafting polymers onto surfaces is a versatile and stable method for
controlling wettability, lubrication, adhesion and surface interactions
\cite{Zhao00}. Brush coatings are utilised for biocompatibilisation or as
antifouling coatings in microfluidic devices. They are employed to stabilise
colloids or, in form of interfacial layers of copolymers, to prevent
coalescence of droplets in polymer blends.  \footnote{Copolymers at interfaces
in blends can laterally move, but the ends of grafted chains in a brush are
immobile.} Dense polymer brushes give rise to fascinating dynamical properties
like a dramatic reduction of friction between two brushes sliding past each
other \cite{Klein94b}, incompletely understood collective dynamics
\cite{Yakubov04}, additional dissipation mechanisms of droplets moving on
brushes due to the the deformability of the soft elastic surface
\cite{Carre96}, and large effective slip lengths \cite{Fetzer05}. 

%previous work
Simulations of dense polymer brushes with an {\em explicit solvent}
\cite{Peters95, Grest96, GrestAdv, Pastorino06} are computationally very
demanding and previous studies of brushes in shear flow \cite{Lai93, Miao96,
Doyle97b, Doyle98, Saphiannikova98} conceived the brush as a porous medium
\cite{Milner91b} utilising Brinkman's equation \cite{Brinkman47} to quantify
the penetration of the solvent flow into the brush. They observed that density
profiles normal to the surface are not strongly affected by shear flow but the
average molecular conformations are tilted towards and stretched along the
direction of the flow.

Isolated grafted chains in shear flow already exhibit intriguing dynamics
\cite{Doyle00, Gerashchenko06, Delgado-Buscalioni06, Winkler06}:  Thermal
fluctuations away from the surface expose the non-grafted chain end to a faster
flow and lead to an elongation of the chain in the flow direction. The
stretched chain rotates back towards the grafting surface and contracts.
Related effects have also been reported for the chain ends in swollen brushes
under strong shear and they have been attributed to the non-linear elasticity
of the macromolecules \cite{Saphiannikova98}.  One hallmark of this dynamic
behaviour of isolated chains are strongly non-Gaussian distribution functions
of angular orientations \cite{Gerashchenko06, Winkler06}. Explicit hydrodynamic
interactions \cite{Delgado-Buscalioni06, Winkler06} appear not to be crucial
for this cyclic tumbling motion.  

%summary
Using Single-Chain-in-Mean-Field (SCMF) simulations and Molecular Dynamics (MD)
simulations of a brush in contact with an explicit solvent of identical chain
molecules we demonstrate that the tumbling motion observed for isolated chains
persist as one increases the grafting density and that the individual motion of
the tethered molecules results in the inversion of the collective flow
direction at the surface. The effect of the brush cannot be described by a
Navier slip condition \cite{Navier1823, Wolynes76d, Bocquet93}.

\section{Model and simulation technique}
%SCMF simulations
%definition of the model
Brush and melt polymers are represented by a bead-spring model with $N=32$
segments per molecule. Harmonic bonding forces ${\bf F}_{\rm b}$ with spring
constant $3 (N-1) k_BT/R_e^2$ act between neighbours along the chain.  $R_e^2$
denotes the mean square end-to-end distance, $k_B$ Boltzmann's constant and $T$
temperature, respectively.  Repulsive interactions between segments are catered
for by penalising fluctuations of the local density by an energy
\cite{Helfand72}, ${\cal H}_{\rm nb} = \frac{\kappa k_BT}{2} \rho_0 \int {\rm
d}^3{\bf r}\; \left( \phi_{\rm b} + \phi_{\rm m} -1 \right)^2$ where $\phi_{\rm b}$
and $\phi_{\rm m}$ denote the normalised density of brush and melt, and $\kappa
N = 50$ characterises the inverse compressibility.  The segment number density,
$\rho_0$, determines the invariant degree of polymerisation, $\bar{\cal N}=
(\rho_0 R_e^3/N)^2=128^2$.
We study the flow of a polymer melt over a brush in a slit pore of width $D_x$.
The lateral extensions are $L_y=L_z$, and periodic boundary conditions are
applied in $y$ and $z$ directions. At $x=0$ and $D_x$, there are repulsive,
structureless walls \footnote{The melt perfectly slips over the bare surface
(without brush). At the considered grafting density, however, the velocity at
the wall vanishes (microscopic stick, see Figs.~\ref{fig:2} and \ref{fig:4}).
Thus, our results do not depend on the microscopic boundary condition (slip or
stick).} that repel segments via a potential \footnote{A repulsive $V_{\rm w}$
impedes the SMC to propose moves that would penetrate the hard walls and be
rejected.} of the form: $V_{\rm w}(\Delta x) = \Lambda_{\rm w} \exp(-\Delta
x^2/2\epsilon_{\rm w}^2)$ with $\Lambda_{\rm w} N= 66.\bar{6}k_BT$,
$\epsilon_{\rm w}=0.15R_e$, and $\Delta x$ being the distance of a segment from
the surface.  The first segment of the brush is grafted at $\Delta x_0=0.1R_e$.  
Couette flow is imposed by moving the grafting points of the brush at the top
surface at fixed velocity in $y$-direction; Poiseuille flow is generated by applying an external
body force, ${\bf F}_{\rm ex}$, to all segments in $y$-direction.

%SCMF simulations
In SCMF simulations \cite{Muller05,Daoulas06b} one considers a large ensemble
of explicit chain configurations that independently evolve in an external
field. This field mimics the effect of the non-bonded interactions $W({\bf r})
= \delta {\cal H}_{\rm nb}/\delta [\rho_0\phi_{\rm b}({\bf r})]+V_{\rm w}$.
Its gradient gives rise to a force ${\bf F}_{\rm nb}= - \nabla W$.
Chain segments evolve via Smart Monte Carlo (SMC) moves \cite{Rossky78,allen86} with
trial displacements 
\begin{equation}
\Delta {\bf r}_{\rm trial} =  \left( \zeta \langle \bar {\bf v}\rangle
+ {\bf F} \right) \Delta t/\zeta + \xi\sqrt{2k_BT \Delta t/\zeta} 
\end{equation}
${\bf F}={\bf F}_{\rm b} + {\bf F}_{\rm nb} + {\bf F}_{\rm ex}$ denotes the
forces acting on a segment, $\zeta$ is the segmental friction,\footnote{The
segmental friction, $\zeta$, is not determined by the intermolecular forces
like in a MD simulation but it is a parameter of the algorithm.} and  $\xi$ is
a Gaussian random number with zero mean and unit variance. We set $\zeta N=1$.
\footnote{For Brownian dynamics the equations of motion are invariant under
$\zeta \to \zeta'=\lambda \zeta$ and $t \to t'=\lambda t$ and we measure time
in units of $\zeta N R_e^2/k_BT$} $\langle \bar {\bf v}({\bf r})\rangle$
denotes the hydrodynamic velocity field.  Chains exhibit Rouse-like dynamics
characteristic for unentangled melts \cite{Doi} and the Weissenberg number is
defined as Wi$\equiv\dot \gamma \zeta N R_e^2 / (3\pi^2 k_BT)$ where $\dot
\gamma$ denotes the shear rate.

The {\em fluctuating, external field}, $W$, is calculated after every time step,
and the densities are constructed by assigning the instantaneous particle
positions to a three-dimensional grid with linear spacing $\Delta L=R_e/6$
\cite{Daoulas06b}. The hydrodynamic velocity, $\langle \bar {\bf v} \rangle$,
however, represents the {\em average flow field} and must not fluctuate. First,
we directly calculate an instantaneous velocity, ${\bf v}_i=\Delta {\bf
r}_i/\Delta t$, of segment $i$ from its explicit displacement \cite{Narayanan06}
during a SMC step (in order to retain spatial resolution) and assign it to the
grid. Then, we add the instantaneous velocities of all segments, average over a
time period, $T$, and normalise by the local density to obtain the average local
velocity $\langle \bar {\bf v}({\bf r}) \rangle$.  This procedure ensure that
the average force $\langle \bar{\bf F} \rangle$ vanishes.  This time averaging
procedure limits the simulation technique to stationary or slowly varying
flows. For the density utilised in our simulations a time interval $T\geq 800
\zeta R_e^2/(Nk_BT)$ is sufficient to eliminate fluctuations of the velocity
field and to yield accurate predictions in equilibrium. 

The computational scheme is similar to dynamic self-consistent field techniques
\cite{Fraaije93,Morita01,Lo05b} or self-consistent Brownian dynamics
simulations \cite{Doyle97b,Saphiannikova98,Narayanan04}. With the SMC algorithm
we systematically approach the Brownian Dynamics limit ($\Delta t \to 0$). In
the present work, we use the time step $\Delta t =0.08 \zeta R_e^2/(Nk_BT)$
which closely mimics Rouse dynamics.  The time step is chosen such that the
self-diffusion coefficient and the viscosity in the bulk are close to the Rouse
limit, $D_{\rm R}=k_BT/(\zeta N)$ and $\eta_{\rm R} R_e/(\zeta N
\sqrt{\bar{\cal N}})=1/36$, respectively. The acceptance rate of SMC moves is
about $83\%$. We have explicitly verified that a ten times smaller time step
yields very similar results.  Moreover, our simulations capture density
fluctuations \cite{Daoulas06b} and, most importantly, we do not approximate the
effect of the brush on the solvent flow by that of a porous medium
\cite{Milner91b,Brinkman47} but use an explicit solvent instead.

%MD simulations
To validate the dynamic SCMF simulations, we additionally perform MD
simulations of a standard bead-spring model \cite{Kremer90, Pastorino06}.
Grafted and melt chains comprise $N=32$ segments interacting via a purely
repulsive Lennard-Jones potential with length and energy scales, $\sigma_{\rm
LJ}$ and $\epsilon_{\rm LJ}$, and cut-off $r_c=\sqrt[6]{2}\sigma_{\rm LJ}$.
Neighbouring segments along a chain are bonded together by a Finitely
Extensible Non-linear Elastic (FENE) potential.  The temperature,
$k_BT/\epsilon_{\rm LJ}=1.68$, is maintained by a DPD thermostat
\cite{Pastorino07}.  The MD simulations correspond to a dense brush, $\sigma
R_e^2 = 1.5 \sqrt{\bar{\cal N}}$, of short chains, $\bar{\cal N} = 6^2$.

\section{Results}
%------------------------------------------------------------------------------
\begin{figure}[!t]
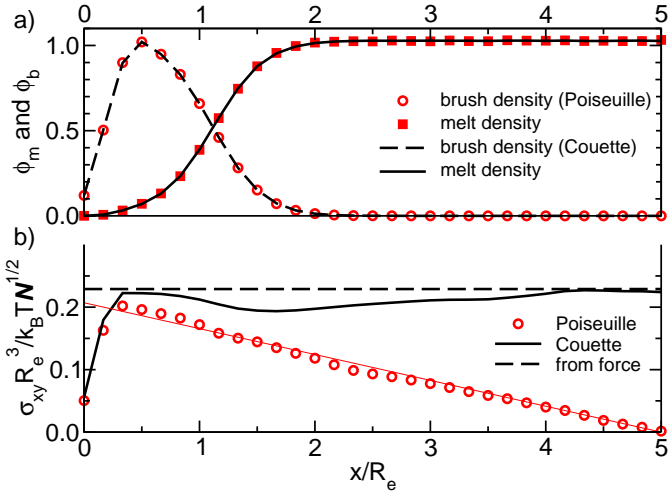

\onefigure[width=\columnwidth]{fig1.eps}
\caption{
a) Density profiles across a slit of width $D_x=10R_e$ and lateral extent
$L_y=L_z=4R_e$ for Poiseuille flow ($F_{\rm ex}N=0.03k_BT/R_e$, symbols) and
Couette flow ($\dot \gamma_{\rm w}\zeta N R_e^2/k_BT = 5$, Wi$=0.247$, lines).
b) Kramer's stress for Poiseuille and Couette flow. The horizontal dashed line
marks the stress obtained from the forces on the grafted ends for Couette flow.
}\label{fig:1}
\end{figure}
%------------------------------------------------------------------------------

%figure 1 a) density profile
In Fig.~\ref{fig:1}a we present the density profile across the slit pore for
$D_x=10R_e$ and grafting density $\sigma R_e^2=\sqrt{\bar {\cal N}}$ for
Couette and Poiseuille flow as obtained by dynamic SCMF simulations. At this
moderate grafting density there is a broad interface between brush and melt
(``wet brush'') and, in agreement with previous studies \cite{Saphiannikova98},
the profile normal to the surface does not depend on flow for small shear
rates. The depletion at the wall stems from the repulsive segment-wall
potential.

\begin{figure}[!t]
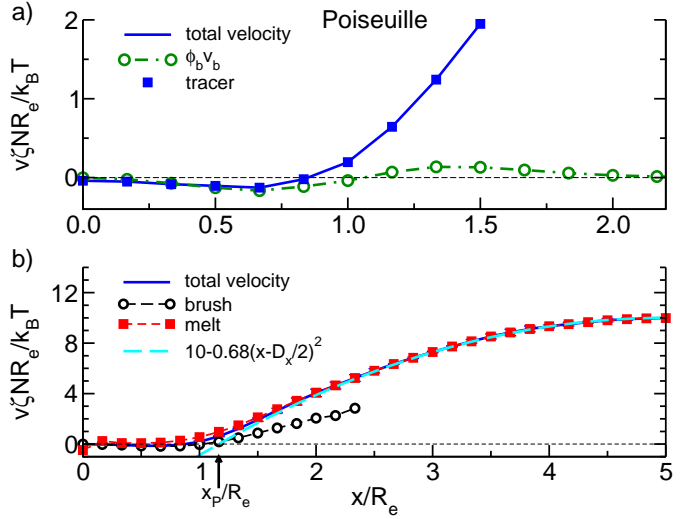

\onefigure[width=\columnwidth]{fig2a.eps}
\caption{
Mass averaged, total velocity profile, $\langle \bar v_y \rangle$, and velocity of
brush and melt segments for Poiseuille flow. Panel (a) highlights the total
velocity at the surface and shows details of $\phi_{\rm b}v_{{\rm b},y}$ for a
system that additionally contains $1\%$ tracer particles (i.e., 6554) which
have the same extension as a monomeric unit and friction coefficient $\zeta =1
$. Note that the velocity profiles of the tracers coincides with $\langle \bar
{\bf v}\rangle$.  
Panel (b) shows the velocity profiles across half of the film. The dashed line
is a fit to macroscopic predictions (cf.~Eqs.~(\ref{eqn:pois})). The arrow marks 
the positions, $x_{\rm P}$, at which the extrapolation of the macroscopic 
hydrodynamic velocity profile vanishes.  
}\label{fig:2}
\end{figure}

\begin{figure}[!t]
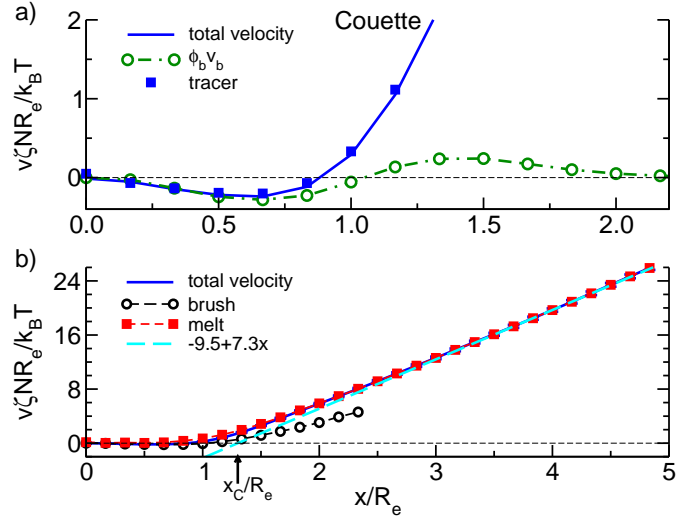

\onefigure[width=\columnwidth]{fig2b.eps}
\caption{
Mass averaged, total velocity profile, $\langle \bar v_y \rangle$, and velocity of
brush and melt segments for Couette flow. Panel (a) highlights the total velocity
at the surface and shows details of $\phi_{\rm b}v_{{\rm b},y}$ for a system
that additionally contains $1\%$ tracer particles (cf.~Fig.~\ref{fig:2}).
Panel (b) shows the velocity profiles across half of the film.  The dashed line
is a fit to macroscopic predictions (cf.~Eqs.~(\ref{eqn:couette})). The arrow
marks the positions, $x_{\rm C}$, at which the extrapolation of the macroscopic
hydrodynamic velocity profile vanishes.
}\label{fig:4}
\end{figure}

%figure 1 b) Kramers stress
Measuring the forces, $F^{\rm graft}_y$, that act on the grafted segments in
Couette flow we obtain the shear stress $\sigma_{xy}R_e^3/k_BT
%\frac{F^{\rm graft}_yR_e^3}{L_yL_zk_BT} 
%= \frac{\eta \dot{\gamma} R_e^3}{k_BT} \simeq \frac{\pi^2}{12} {\rm Wi} \sqrt{\bar{\cal N}} 
=29.3$  
for Couette flow.  Alternatively, we can estimate the intramolecular stress of
the melt at the centre of the film via the mean-field approximation (Kramer's
formula \cite{Doi} \footnote{We utilise the same spatial assignment of the
stress as for the density but other schemes [e.g., J. H. Irving and J. G.
Kirkwood, {\it J. Chem.  Phys.}, {\bf 18} (1950) 817.] that are locally more
accurate can be employed.}) $\sigma_{xy}R_e^3/(k_BT\sqrt{\bar {\cal N}}) =
(N-1) (\phi_{\rm b}+\phi_{\rm m})\frac{\langle b_x b_y
\rangle}{R_e^2/[3(N-1)]}$ where $b_x$ and $b_y$ denote the distance between
bonded segments perpendicular to the surfaces and along the shear direction,
respectively. This estimate is shown in Fig.~\ref{fig:1}b for Poiseuille and
Couette flow.  The stress in Couette geometry agrees well with the result
obtained from the force on the grafted segments. From the stress at the centre
of the film and the velocity gradient (cf.~Figs.~\ref{fig:2} and \ref{fig:4}) 
we can estimate the dimensionless viscosity, 
$\eta R_e/(\zeta N \sqrt{\bar{\cal N}})=0.030$ and
$0.031$ for Poiseuille and Couette flow, respectively.

%------------------------------------------------------------------------------
%------------------------------------------------------------------------------

Fig.~\ref{fig:2} presents the velocity across the film for Poiseuille flow. The profile
at the centre is parabolic, 
\begin{equation}
v_{{\rm hydro},y}(x) = \frac{\rho_0F_{\rm ex}}{2\eta} \left( x-x_{\rm P}\right) \left( D_x-x_{\rm P}-x\right)
\label{eqn:pois}
\end{equation}
as expected from macroscopic hydrodynamics. $x_{\rm P}$ denotes the position at
which this parabolic profile extrapolates to zero.  The flow penetrates into
the brush \cite{Milner91b} and details of the flow in the vicinity of the
surface are shown in the panel (a) of Fig.~\ref{fig:2}.  Unexpectedly, the
hydrodynamic velocity profile in the vicinity of the grafting surface becomes
negative, {\em i.e.}, the flow direction is {\em opposite} to the driving force
$F_{\rm ex}$. This inversion of the collective flow direction at the grafting
surface is a consequence of the individual motion of grafted chains, i.e., the
tumbling motion of grafted chains under shear \cite{Doyle00,Saphiannikova98}.

%------------------------------------------------------------------------------
\begin{figure}[!t]
\onefigure[width=\columnwidth]{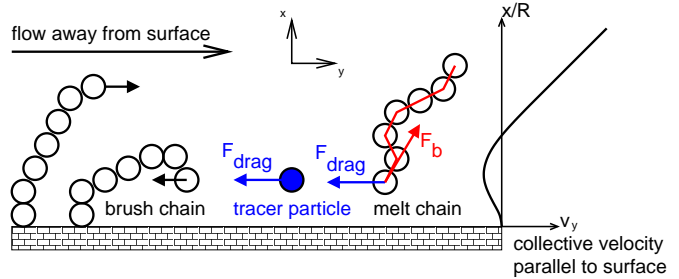}
\caption{Sketch of the molecular motion of the grafted molecules of the brush,
the free chains of the melt and the smaller tracer particles.  On the left,
brush molecules in two different stages of their tumbling motion are depicted
and the velocity of the end-segment is highlighted by an arrow. This graph
illustrates that the collective velocity of the brush segments {\em at a
specific distance from the surface} need not to vanish although the
time-averaged velocity of each segment of grafted chains is zero. In the
middle, the forces that act on the tracer particle and a segment of a free
chain near the surface are illustrated. On the right hand side, the collective,
hydrodynamic velocity profile, $\langle \bar {\bf v} \rangle$, of all segments
is shown.  
} \label{fig:3} 
\end{figure}
%------------------------------------------------------------------------------

To rationalise the observed inversion of flow at the surface, it is important
to distinguish between the time-averaged velocity of individual segments and
the position-dependent, hydrodynamic flow,  $\langle \bar {\bf v}({\bf r})
\rangle$.  In the steady state, the time-averaged velocity of each individual
segment of the grafted chains must vanish because they are irreversibly grafted
to the substrate. This property implies $\int {\rm d}x\; \phi_{{\rm b}}(x)
v_{{\rm b},y}(x) = 0$, which is obeyed in our simulations as shown in
Fig.~\ref{fig:2}a.  The drag force, exerted by the $x$-dependent, hydrodynamic
flow field, $\langle \bar v_y\rangle(x)$, couples the motion in the parallel
and perpendicular directions and gives rise to a correlation between the
position, $x$, of a brush segment and its velocity, $v_y$. A brush segment that
is farther away from the grafting surface experiences a larger drag force and
has, on average, a larger parallel velocity, $v_{{\rm b},y}$, than the same
segment when it is located closer to the grafting surface and exposed to a
smaller (inverted) drag. This effect is illustrated in Fig.~\ref{fig:3}.
In Fig.~\ref{fig:2}b, one clearly observes that the very few segments of the
brush that extend far into the melt, $x/R_{\rm e}>2$, are dragged along with
the flow and have a large positive velocity ($v_{{\rm b},y}(x) \gtrsim
\frac{1}{2}\langle \bar v_y\rangle(x)$).  Since they are grafted, they must
compensate this by a negative velocity when they are closer to the grafting
surfaces. This mechanism results in the cyclic motion of the individual
molecules. Our simulations show that this cyclic motion is not necessarily a
consequence of the non-linear elasticity of the chain molecules
\cite{Saphiannikova98} and that it also persists in moderately dense brushes.
Averaging the velocity of all brush segments in the vicinity of the surface, we
obtain a negative velocity and, since the density of the melt chains is low at
the surface, the negative velocity inside the brush is sufficient to invert the
{\em total, mass averaged velocity profile} $\langle \bar {\bf v} \rangle$
(cf.~Fig.~\ref{fig:2}a).

For the choice of parameters, the velocity profile of the free chains of the
melt does not exhibit an inversion of the flow direction at the surface. Since
the free chains only weakly penetrate the brush, the larger part of the
extended molecule is exposed to flow in the positive direction at the centre
even if a single segment is located close to the surface. In this case, the
drag force on this near-surface segment is compensated by the bonding force (as
sketched in the middle of Fig.~\ref{fig:3}) and no inversion of the flow is
observed.
To observe the inversion of the flow direction of freely moving particles at
the surface, we have added $1\%$ of monomeric tracer particles (i.e., 6554
beads), which have identical size than the polymer segments and friction
coefficient, $\zeta=1$. In Fig.~\ref{fig:2}a, we demonstrate that the velocity
profile of these free tracer particles exhibits flow inversion at the surface
and that such particles potentially can be used to experimentally detect the
inversion of the flow direction at the surface.

We expect that the inversion of the velocity is characteristic for intermediate
grafting densities. At very small grafting densities, the flow at the surface
is dominated by the velocity of the solvent and the cyclic motion of the
grafted chains will not be sufficient to invert $\langle \bar {\bf v} \rangle$.
Contrary, if the brush was strongly stretched, the width of the brush-melt
interface would be narrow, the height fluctuations of brush segments would be
small and -- in the Lennard-Jones model and experiments -- topological
constraints would become more important. These effects would shift the
inversion zone away from the grafting surface and tend to reduce the inversion
of the flow.

%------------------------------------------------------------------------------
%------------------------------------------------------------------------------

To verify that the inversion of velocity profile at the brush-coated surface is
independent from the way the flow is generated we have studied Couette flow,
moving the grafted chain ends at the top surface with constant velocity $\dot
\gamma_{\rm w} D_x$. The velocity profile across the pore is shown in
Fig.~\ref{fig:4}. \footnote{Only the bottom half of the slit is shown in
Fig.~\ref{fig:4} Due to Galilean invariance, the flow velocity at the top wall
is positive and segments at the top wall move in the positive direction with a
higher velocity than the top wall.}  The linear behaviour at the centre
corresponds to macroscopic hydrodynamics
\begin{equation}
v_{{\rm hydro},y}(x) = \dot{\gamma}\left(x-x_{\rm C}\right)
\label{eqn:couette}
\end{equation}
but at the surface the flow direction is inverted. $x_{\rm C}$ denotes the
location where this macroscopic linear velocity profile vanishes. If we used
this as the definition of an effective slip length, the slip length would be
negative \cite{Pastorino06}.  
%\footnote{Alternative definitions which specify both -- slip length and effective position at which the hydrodynamic boundary condition is to be applied -- are possible. In this case, $x_{\rm C}$ would approximately correspond to the position of the boundary and the slip length would be small.} 

%------------------------------------------------------------------------------
\begin{figure}[!t]
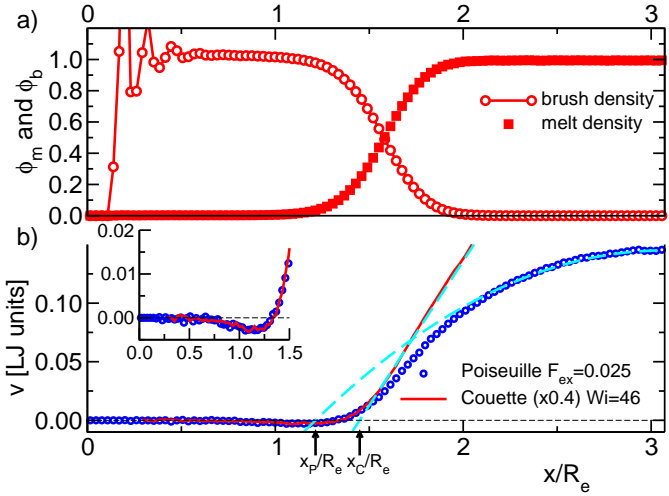

\onefigure[width=\columnwidth]{fig3.eps}
\caption{
a) Density profile of the brush-melt interface obtained from the MD simulations using a bead-spring model ($N=32, R_e=6.51\sigma_{\rm LJ}$,
$\sigma R_e^2=8.9, \rho_0\sigma_{\rm LJ}^3=0.69, \sqrt{\bar{\cal N}}=5.95), k_BT/\epsilon_{\rm LJ}=1.68$).
b) Mass averaged velocity profile $\langle \bar v_y \rangle$ for Poiseuille flow with $F_{\rm ex}\sigma_{\rm LJ}=0.025\epsilon_{\rm LJ}$ and Couette flow with Wi$=46$. The data for Couette flow have been scaled by a factor $0.4$. The arrows mark $x_{\rm P}$ and $x_{\rm C}$.
The inset highlights the velocity at the surface.  
}\label{fig:5}
\end{figure}
%------------------------------------------------------------------------------

%figure 3: MD simulations
To validate this observation, we additionally perform MD simulations of a standard Lennard-Jones bead-spring model.
Fig.~\ref{fig:5}a presents the density profiles of brush and melt segments
while panel b presents the total velocity profiles for Poiseuille and Couette
flow. The inversion of the flow direction inside the brush is confirmed.
Importantly, the MD simulations do not invoke any approximation: They do not
ignore inertia effects and do not replace intermolecular interactions by quasi-instantaneous external fields \cite{Daoulas06b}.
They capture the local packing of a dense polymeric
fluid (see density oscillations in Fig.~\ref{fig:5}a), the finite extensibility
of bonds and, importantly, the non-crossability of the chain molecules.

%hydrodynamic boundary condition
Utilising the simulation results of Poiseuille and Couette flow we try to
extract the parameters of the hydrodynamic boundary condition \cite{Bocquet93}
that serves to match the detailed microscopic velocity profile in the vicinity
of the brush-coated surface with macroscopic hydrodynamics, ${\bf v}_{\rm hydro}$.
The Navier slip condition \cite{Navier1823} assumes that the
viscous stress $\sigma^{\rm visc}_{xy} = \eta \frac{{\rm d}v_{{\rm
hydro},y}}{{\rm d}x}$ calculated from the macroscopic hydrodynamic velocity profile,
${\bf v}_{\rm hydro}$, equals the friction stress $\sigma^{\rm fric}_{xy} =
\frac{\eta}{\delta_{\rm b}} v_{{\rm hydro},y}$ at a distance, $x_{\rm b}$.  $\delta_{\rm b}$ denotes
the slip length and $x_{\rm b}$ the effective position of the hydrodynamic
boundary. Both parameters are assumed to be materials properties that
characterise the coated surface independently from the strength and type of flow.
From Eqs.~(\ref{eqn:pois}) and (\ref{eqn:couette}) one obtains $\delta_{\rm b}
= \sqrt{(x_{\rm P}-x_{\rm C})(D-x_{\rm C}-x_{\rm P})}$ and $x_{\rm b}=x_{\rm
C}+\delta_{\rm b}$. Thus, a necessary condition for simultaneously describing
both types of flows by the same parameters, $x_{\rm b}$ and $\delta_{\rm b}$,
is $x_{\rm P} \geq x_{\rm C}$ which is neither fulfilled by the profiles
extracted from the SCMF simulations with $D_x=10R_e$ (see Figs.~\ref{fig:2} and \ref{fig:4}) nor
by those obtained from MD simulations of the bead-spring model with
$D_x=6.14R_e$ (cf.~Fig.~\ref{fig:5}).

%figure 4: tumbling statistics
%------------------------------------------------------------------------------
\begin{figure}[!t]
\onefigure[width=\columnwidth]{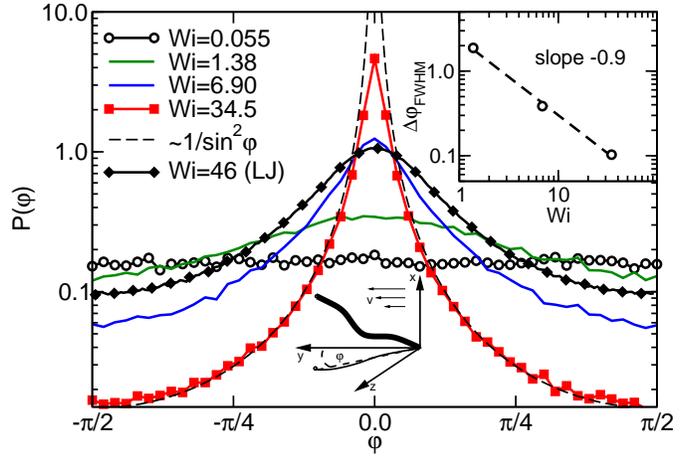}
\caption{Probability distribution of the angle, $\varphi$ (see lower sketch).
For moderate shear, the distribution is sharply peaked around $\varphi=0$ and
its wings are compatible with $P(\varphi) \sim 1/\sin^2(\varphi)$ (dashed
line).  SCMF simulation data are obtained for $D_x=6R_e$ and $L_y=L_z=2R_e$.
The time step $\Delta t$ for the highest shear rate, Wi$=34.5$, has been
decreased by a factor $0.1$. The MD simulation data of the bead-spring model
with Wi$=46$ are presented by filled symbols.  The upper inset presents the
width of the distribution as a function of the Weissenberg number.  }
\label{fig:6} 
\end{figure}
%------------------------------------------------------------------------------

Finally, in Fig.~\ref{fig:6} we analyse the fluctuations of the angle, $\varphi$,
between the projection of the chain's end-to-end vector on the grafting surface
and the flow direction (see inset) for different shear rates. Similar to the
behaviour of isolated chains \cite{Gerashchenko06, Winkler06}, the wings of the
distribution for moderate Weissenberg numbers are compatible with $P(\varphi)
\sim 1/\sin^2 \varphi$. This non-Gaussianity of the distribution is
corroborated by the MD results. At comparable Weissenberg number, the effect is
smaller for the Lennard-Jones brush because of the higher grafting density and
non-crossability.  The full width at half maximum $\Delta \varphi_{\rm FWHM}$
of the distribution is compatible with $\varphi_{\rm FWHM} \sim {\rm
Wi}^{-0.9}$ in agreement the corresponding predictions for isolated chains and
low shear rates.

\section{Conclusion}
In summary, dynamic SCMF and MD simulations of polymer brushes in an explicit
solvent show that the motion of chains in a dense brush is similar to the
cyclic tumbling of isolated grafted chain in shear flow \cite{Doyle00}.  We
find that brush chains are not uniformly tilted by the flow but the orientation
of the molecular axis with the flow direction is characterised by a
non-Gaussian distribution \cite{Gerashchenko06, Winkler06}.  The individual
tumbling motion of the brush molecules leads to an inversion of the total,
collective flow velocity at the grafting surface.  It would be interesting to
quantitatively compare simulations with an implicit solvent using Brinkman's
equation and our simulations with an explicit solvent to assess to what extent
the brush can be described as a static porous medium.  The simulation also
provide microscopic insights into the mechanisms that dictate the hydrodynamic
boundary condition. The latter is important for controlling flow in
microfluidic devices.  The cyclic tumbling motion of the individual brush
molecules could also potentially serve to separate particles with different
surface affinities by flow or to promote mixing close to the grafting surface.

\acknowledgments
We thank M.P.~Allen, K.Ch.~Daoulas, G.H.~Fredrickson, A.~Milchev, S.~Minko,
and C.~Servantie for inspiring discussions.  Financial support was provided by
the priority program ``Nano- and Microfluidics'' MU1674/3-2, the INFLUS
project, the DAAD/SECYT, and ANPCYT (PME 2003, PICT 2005). Computing time at
NIC J{\"u}lich, HLRN Hannover, and GWDG is acknowledged.

\bibliographystyle{epl}
\bibliography{bibtex}
\end{document}